\documentstyle[aps,prl,floats,epsf]{revtex}

\begin{document}

\twocolumn[\hsize\textwidth\columnwidth\hsize\csname
@twocolumnfalse\endcsname

\title{Ensemble Density-Functional Theory for ab-initio Molecular Dynamics
of Metals and Finite-Temperature Insulators}
\author{Nicola Marzari,$^{1,2}$ David Vanderbilt,$^1$ and M.\ C.\ Payne$^2$}
\address{$^1$Department of Physics and Astronomy, Rutgers University, 
Piscataway, New Jersey 08855-0849 \\
$^2$Cavendish Laboratory (TCM), University of Cambridge, Madingley
Road, Cambridge CB3 0HE, England}

\date{March 8, 1997}

\maketitle

\begin{abstract}
A new method is presented for performing 
first-principles molecular-dynamics simulations of
systems with variable occupancies.
We adopt a matrix representation for the one-particle 
statistical operator $\hat{\Gamma}$, to introduce a ``projected''
free energy functional $G$ that depends on the Kohn-Sham orbitals only
and that is invariant under their unitary transformations.
The Liouville equation ${[}\hat{\Gamma},\hat{\rm H}{]}=0$ is always satisfied, 
guaranteeing a very efficient and stable variational 
minimization algorithm that can be extended to non-conventional
entropic formulations or fictitious thermal distributions.
\end{abstract}
\vskip1pc]

\narrowtext

In recent years, the range of problems that can be studied with
quantitative accuracy using the methods of computational solid
state physics has expanded dramatically. It is now possible to
calculate many materials properties with
an accuracy that is often comparable to that of experiments.
This degree of confidence is based on the fundamental quantum-mechanical
treatment offered by density-functional theory (DFT)\cite{dft-lda},
coupled with the availability of increasingly powerful computers
and with the development of algorithms tuned towards optimal
performance\cite{cp-cg}.

The application of these methods and techniques to metallic
systems has nonetheless encountered several difficulties,
that have made progress slower than for the case of semiconductors
and insulators.  The 
discontinuous variation of the orbital occupancies across the
Brillouin zone (BZ) makes the occupation numbers 
rather ill-conditioned variables, and the self-consistent solution of
the screening problem can suffer from several instabilities.
The absence of a gap in the energy spectrum
and the requirement of an exact
diagonalization for the Hamiltonian matrix everywhere in the BZ
(in order to assign the occupation numbers) 
introduce ``slow frequencies'' in the evolution of the orbitals
towards the ground state and
preclude the straightforward
extension to metals of algorithms which performed well for insulators.  
Smearing the Fermi surface with a 
finite electronic temperature\cite{smearing} allows for an
improved BZ sampling, but only partially alleviates the
problems alluded to above.

In this Letter, we introduce a new approach which solves many of
these problems in a natural way, and which provides a general
and efficient framework for obtaining the ground state of a
Kohn-Sham Hamiltonian at a finite electronic temperature. The
typical context is the Mermin formulation for the Fermi-Dirac
statistics\cite{mermin}, but the method also applies when
generalized entropic functionals are introduced\cite{genentropy},
as it is often the case for metallic systems.
Other applications include DFT studies of 
insulators or semiconductors with thermally excited states\cite{alavi},
and fractional quantum Hall states\cite{fqhe}.
The language of ensemble-DFT\cite{ensemble} is used, and an 
orbital-based variational algorithm for the minimization to 
the ground state is developed and implemented.  
Dramatic improvements are obtained
in the convergence of the energies and especially of the
Hellmann-Feynman forces.

Within ensemble-DFT, the Helmholtz free energy functional 
at a temperature $T$ and for an $N$-re\-pre\-sen\-ta\-ble 
charge density $n({\bf r})$ in an external potential $V$ is
$A_V[n({\bf r})]=F_T[n({\bf r})]+\int V({\bf r}) n({\bf r}) d{\bf r}$,
where $F_T$ is the finite-temperature Mermin-Hohenberg-Kohn 
functional\cite{mermin}.
The charge density $n_0({\bf r})$ that minimizes $A_V$ is the ground-state 
charge density, and $A_V[n_0({\bf r})]$ is the free energy 
of the electronic system.
A Kohn-Sham-like mapping onto non-interacting electrons 
leads to a decomposition of the functional $F_T$ into
explicit terms (the non-interacting kinetic
energy, the classical electrostatical energy, and the entropic contribution)
plus the unknown exchange-correlation functional $E_{T,\rm xc}$, for which
we take here the local density approximation\cite{dft-lda}.

A key assumption is made by adopting a matrix representation $f_{ij}$, in
the basis of the orbitals, for the
one-particle effective statistical operator $\hat{\Gamma}$;
the charge density is correspondingly written as
\begin{equation} \label{eq:nfree}
  n({\bf r})\,=\,\sum_{ij} f_{ji}\, \psi_i^* ({\bf r}) \psi_j({\bf r}) \;\;.
\end{equation}
Here the $\{\psi_i\}$ are orthonormal single-particle Kohn-Sham orbitals,
the sum extends in principle over all the states, and
the matrix $f_{ij}$ is constrained to have trace equal to the number of
electrons and eigenvalues bounded between $0$ and $1$.
We can then write in all generality the functional 
$A$ to be minimized as
\begin{equation} 
\label{eq:afree}
A\,[\,T\,;\{\psi_i\}\,,\,\{f_{ij}\}\,]\,=
\end{equation}
$$\sum_{ij}\,f_{ji}\,\langle\psi_i\vert \hat{\rm T}+\hat{\rm V}_{\rm ext} \vert\psi_j\rangle
\,+\,E_{\rm Hxc}[n]\,-\,T S[\,\{f_{ij}\}\,] \;\;;$$
the Hartree and exchange-correlation terms, which depend on the
charge density, have been grouped together. 
The entropic term is taken to be a function of the eigenvalues of 
${\bf f}$; in the Fermi-Dirac case, it is $S[\,\{f_{ij}\}\,]={\rm tr}\, s({\bf f})$, where
$s$ is $f \ln f + (1-f) \ln (1-f)$.
In most applications, the external potential $V_{\rm ext}$ is 
generated by an array of non-local ionic pseudopotentials.
The free energy functional defined in Eq.~(\ref{eq:afree}) is 
in the form of traces of operators, and so it is covariant 
under a change of representation (i.e., for a unitary transformation 
${\bf U}$ of the orbitals $\{\psi_i\}$); this can be verified by
letting $ {\bf f} \rightarrow {\bf f}^\prime={\bf U}\,{\bf f}\,{\bf U}^\dagger $ and 
$ \vert\psi_j\rangle \rightarrow \vert\psi_j\rangle^\prime = 
\sum_{m} U_{jm}^* \psi_m $.
The covariance of the free energy functional $A$ allows
for the definition of a new, {\em projected functional G}, 
that depends only on the orbitals $\{\psi_i\}$:
\begin{equation} \label{eq:g}
  G\ [\,T\,;\{\psi_i\}\,] := 
  \min_{\{f_{ij}\}}  A\,[\,T\,;\{\psi_i\}\,,\,\{f_{ij}\}\,]\;\;.
\end{equation}
$G$ is {\em invariant} under
any unitary transformation of the $\{\psi_i\}$: 
the transformed orbitals cannot lead to a different
value for $G$, by virtue of the covariance of $A$.

The projected functional $G$ represents
a much better conditioned choice
than the original free energy $A$ when it is used
in minimization algorithms that are based on the orbital propagation towards
self-consistency, as is the case in the Car-Parrinello or 
conjugate gradients methods\cite{cp-cg}. The reasons are several, 
albeit related.
(1) The functional $G$ no longer depends on the 
occupancies of the orbitals or on their 
unitary transformations (``rotations'') in the occupied subspace.
These are ill-conditioned, non-local degrees of 
freedom, with the added non-linear constraint of charge normalization.
(2) The $f_{ij}$ in this formalism have become dependent variables,
implicitly defined by the minimization in Eq.~(\ref{eq:g}), and this dependency
does not enter into the calculation of the functional derivatives ${\delta
G}/{\delta \psi_i^*}$, since the contributions $({\partial
G}/{\partial f_{kl}})({\partial f_{kl}}/{\partial \psi_i^*})$ are
zero because of the minimum condition. 
(3) The occupancies of the orbitals and their rotations in the occupied
subspace are now consistently considered as part of the same problem,
namely that of finding the ground-state statistical operator, and 
not---as it is usually
done---as two independent degrees of 
freedom. The evolution of the $f_{ij}$ turn out to be largely decoupled 
from the problem of updating the orbitals in the subspace orthogonal to 
the occupied subspace.
(4) The expensive and inefficient evolution for the orbital
rotations is now shifted to the matrix ${\bf f}$; this implies that the
associated slow frequencies in the evolution of $A$ have now been
compressed to zero by the minimization condition.
Subspace alignment\cite{apj} between subsequent
orbital updates is also automatically enforced.

This formulation naturally separates the evolution of the orbitals $\{\psi_i\}$
from that of the $f_{ij}$: the orbitals get updated in an outer loop 
that minimizes $G$ to selfconsistency, and after every update
an inner loop on the $f_{ij}$ minimizes $A$ {\it at fixed orbitals} $\{\psi_i\}$.
In other words, the $f_{ij}$ are projected via the inner loop
onto the $G$ surface, where ${\bf f}$ (or $\hat{\Gamma}$, of
which ${\bf f}$ is the matrix representation in the orbital basis) commutes
with the non-selfconsistent Hamiltonian. The minimization of $G$ is
then freed from all constraints but the orthonormality of the $\{\psi_i\}$,
and the rotations of the orbitals do not play any role.

The commuting relationship of $\hat{\Gamma}$ and $\hat{\rm H}$ can be made
explicit from the minimum condition. Let us define
$$ h_{ij}\,=\,\langle\psi_i\vert \hat{\rm T} + \hat{\rm V}_{\rm ext} \vert\psi_j\rangle \;\;,\;\;
 V_{ij}^{[n]}\,=\,\langle\psi_i\vert \hat{\rm V}_{\rm Hxc}^{[n]} \vert\psi_j\rangle $$
for the matrix elements of the Hamiltonian (the superscript $[n]$ is
a reminder that the potential ${\rm V}_{\rm Hxc}^{[n]}$ depends self-consistently
on the charge density); the minimum condition that defines $G$ implies
\begin{eqnarray}
{\delta A\over\delta f_{ji}}\,&=&\,0\,=\,h_{ij}+{\delta E_{\rm Hxc}\over\delta f_{ji}}
  -T{\delta S\over\delta f_{ji}} -\mu\,\delta_{ij}
\nonumber\\
&=&\,h_{ij}+\int\,d{\bf r}\,{\delta E_{\rm Hxc}\over\delta n({\bf r})}\,
    {\delta n({\bf r})\over\delta f_{ji}} -T\,[s^\prime({\bf f})]_{ij} -\mu\,\delta_{ij}
\nonumber\\
&=&\,h_{ij}+V_{ij}^{[n]} -T\,[s^\prime({\bf f})]_{ij} -\mu\,\delta_{ij}\ \;\;.
\label{eq:grad_a}
\end{eqnarray}
The constraint of charge conservation ${\rm tr}\, {{\bf f}} = N$
is taken into account with the introduction of the Lagrange
multiplier $\mu$, and the notation $\,[s^\prime({\bf f})]_{ij}$ is used in place of
$d\,{\rm tr}\, s({\bf f})\,/df_{ji}\,$.
The stationary condition in (\ref{eq:grad_a}) is thus
\begin{equation}
h_{ij}+V_{ij}^{[n]}-T\,[s^\prime({\bf f})]_{ij}\,=\,\mu\,\delta_{ij} \; \; .
\label{eq:liouville}
\end{equation}
It follows that $f_{ij}$ and the Hamiltonian $h_{ij}+V^{[n]}_{ij}$
are diagonalized by the same unitary rotation, at fixed orbitals,
and thus represent ``commuting'' operators; the non-selfconsistent Liouville 
equation $[\hat{\Gamma},\hat{\rm H}]=0$ is satisfied.
The relation (\ref{eq:liouville}) does not imply
that $h_{ij}+V^{[n]}_{ij}$ and $f_{ij}$ are diagonal,
but just that there is a common transformation that diagonalizes
both; the formalism per se
is not linked to a preferred diagonal representation.

The inner loop for the update of the occupation matrix $f_{ij}$ is carried
out at fixed orbitals, and so it does not require the
calculation of new matrix elements for the kinetic energy operator or the
non-local pseudopotential, and there are no orthogonalizations involved. 
The Fourier transforms of the $\{\psi_i\}$ can also be eliminated
by storing their real-space representation.
We have chosen for the $f_{ij}$ 
an iterative minimization that has a simple and appealing
rationale: if the problem were not self-consistent,
the solution for the equilibrium $f_{ij}$ would be found by 
straightforwardly
diagonalizing the Hamiltonian matrix, calculating from its eigenvalues
the thermal distribution of the occupation numbers,
and rotating them back into the current orbital representation. 
Since the problem is self-consistent, this will not be the actual
solution, but
we use it as a search direction for a direct line minimization
in the multidimensional space of the $f_{ij}$.
The procedure is organized as follows. The matrix $h_{ij}$, with the
kinetic-energy and non-local contributions, is determined
once for all before entering the inner loop.
The updated charge density (let us assume that
the $m^{th}$ iteration in the inner loop is taking place) is calculated as
\begin{equation}
\label{eq:newn}
  n^{(m)}({\bf r})\,=\,\sum_{ij} f_{ji}^{(m)}\, \psi_i^* ({\bf r}) \psi_j({\bf r}) \;\;.
\end{equation}
The Hartree and exchange-correlation energy 
$E_{\rm Hxc}^{(m)}$ and potential $V_{\rm Hxc}^{(m)}({\bf r})$ are then calculated, and
the matrix representation $V_{ij}^{(m)}$ is constructed.
The entropy $S^{(m)}$ is also computed, following a diagonalization of ${\bf f}$,
\begin{equation}
f_{ij}^{(m)}=\sum_l Y_{il}^{(m)\dagger} f_l^{(m)} Y_{lj}^{(m)} \; \; .
\end{equation}
Actually, in a traditional 
plane-wave approach the charge density (\ref{eq:newn})
is calculated more efficiently in this representation in which the $f_{ij}$ 
are diagonal, since a temporary rotation of the orbitals
can then be performed on their more compact reciprocal-space representation.
The Hamiltonian matrix is then updated using the new local terms,
and diagonalized as
\begin{equation}
H_{ij}^{(m)}= h_{ij}+V_{ij}^{(m)}=\sum_l Z_{il}^{(m)\dagger} 
\epsilon_l^{(m)} Z_{lj}^{(m)} \;\;.
\end{equation}
The non-self-consistent minimum for ${\bf f}$ would now be
\begin{equation}
\widetilde{f}_{ij}^{(m)} = \sum_l Z_{il}^{(m)\dagger} 
f_T(\epsilon_l^{(m)}-\mu) Z_{lj}^{(m)} \;\;,
\end{equation}
where $f_T$ is the (Fermi-Dirac) thermal distribution.
We choose this as our search direction in the $f_{ij}$ space,
and a full line minimization is performed along the 
multidimensional segment 
$ {\bf f}_\beta^{(m+1)}={\bf f}^{(m)}+\beta\,\Delta{\bf f}^{(m)} $,
where $\Delta{\bf f}^{(m)}=\widetilde{\bf f}^{(m)}-{\bf f}^{(m)}$.
Note that $\beta$ parametrizes an {\it unconstrained} search,
since ${\rm tr}\,{\bf f}_\beta=N$
at the end-points and thus, by linearity, at all $\beta$.
The search direction is determined by the eigenvalues of the 
non-self-consistent
Hamiltonian, attracting the $f_{ij}$ towards the representation
in which they commute with the Hamiltonian and towards the thermal
equilibrium values that they would assume for this Hamiltonian.
Since the search direction is determined
by the eigenvalues/eigenvectors of $\hat{\rm H}$, and not 
from the occupation numbers, this current formalism can also be 
applied when generalized entropic functionals
are defined, or when non-monotonic thermal distributions are 
introduced\cite{genentropy}.

The direct minimization proceeds by calculating the free energy and its 
derivative along the search line 
at the two end-points $\beta=0$ and $\beta=1$, {\it taking 
into account the self-consistent variations in the charge density}
and thus in the Kohn-Sham Hamiltonian.
The line derivative $A^\prime$ is $\sum_{ij} \Delta f^{(m)}_{ji}\, ({\delta
A}/{\delta f_{ji}})$, where
$$ \left.\frac{\delta A}{\delta f_{ji}}\right|_{\beta=0}\,=\,
\left[\, h_{ij} + V_{ij}^{(m)} -T \sum_l Y_{il}^{(m)\dagger} 
s^\prime(f_l^{(m)}) Y_{lj}^{(m)} \right] ;  $$
$A^\prime(\beta=0)$ is always smaller than $0$, and so the iterative update of ${\bf f}$ 
takes place in a strictly variational fashion.
We then calculate the charge density $\widetilde{n}^{(m)}({\bf r})$, 
the matrix elements $\widetilde V_{ij}^{(m)}$,
and the free energy $\widetilde{A}$ 
corresponding to $\beta=1$, together with the
line derivative $A^\prime(\beta=1)$ via
$$ \left.\frac{\delta A}{\delta f_{ji}}\right|_{\beta=1}\,=\,
\left[\, h_{ij} + \widetilde V_{ij}^{(m)} -T
    \sum_l Z_{il}^{(m)\dagger} s^\prime(\widetilde f_l^{(m)}) Z_{lj}^{(m)} \right] . $$
Since the kinetic energy and the pseudopotential 
contributions are exactly linear along the search direction
(only the prefactors $f_{ij}$ change), and
the Hartree energy is quadratic, 
while the remaining exchange-correlation and entropic terms are
very well-behaved, a cubic or a parabolic interpolation 
locates the value of $\beta$ corresponding to the minimum with very high
accuracy.
More importantly, the choice of a direct minimization for 
${\bf f}$ along a linear search
implies that {\it level-crossing instabilities are completely eliminated},
even in the limit of zero temperature and/or in the absence of an entropic 
term. In practice, we find that 
one or two iterations in the inner loop are the 
optimal choice even for large systems (e.g., the
diffusion of an adatom on a slab of 144 atoms\cite{nicola}), 
since we also need self-consistency with respect to $\{\psi_i\}$.

$G$ can be minimized very efficiently with a direct all-bands 
conjugate-gradients method; however, 
since it has a much broader spectrum for its eigenvalues than in an
insulating case, it exhibits a slower convergence.
This is essentially due to the occupancies of the higher bands 
being close to
zero. To solve this problem, we have resorted to a preconditioning
strategy: we choose a set of scaled variables in which the
functional has a more compressed spectrum, and the search
directions are calculated in this new metric.
In the diagonal representation the total energy around
the minimum is a quadratic form $\sum_{i,n} f_i \epsilon_i
c_{i,n}^2$ ($c_{i,n}$ is the expansion coefficient of $\psi_i$ in
the $n$-th element of the basis set); if the steepest-descent
directions are chosen according to scaled variables 
$\widetilde{c}_{i}= \sqrt{f_i}\,c_{i,n}$, the preconditioned 
steepest-descent directions for the original variables 
become\cite{gillan}
\begin{equation}
\label{eq:prec}
-\frac{\delta G}{\delta\psi_i^*}\,\rightarrow\,
-\frac{1}{f_i}\,\frac{\delta G}{\delta 
\psi_i^*}\,=\,-\,\hat{\rm H}\,\psi_i\;\;.
\end{equation}
With some degree of overcorrection\cite{prec}, these can also be used 
to construct conjugate directions.
A generalization to our case is obtained
by calculating the steepest-descent directions ${\bf g}_i$ of
$G$, passing them into the diagonal representation
where they can be preconditioned as in (\ref{eq:prec}), 
and transforming them back in the initial representation. The 
steepest-descent directions are
\begin{equation}
\label{eq:g_i}
{\bf g}_i\,=\,-\frac{\delta G}{\delta \psi_i^*}\,=\,-\sum_j f_{ji} 
\hat{\rm H} \vert\psi_j\rangle\; ; \; {\bf g}_i^\prime\,=\,
-f^\prime_{ii} \hat{\rm H} |\psi_i^\prime\rangle \;,
\end{equation}
where the primed term refers to the diagonal representation 
($ {\bf f}^\prime =f^\prime_{ii}\,\delta_{ij}= {\bf U}\,{\bf f}\,{\bf U}^\dagger$).
The preconditioned steepest descents ${\bf G}^\prime_i$ and ${\bf G}_i$ are
thus 
\begin{equation}
 {\bf G}^\prime_i\,=\,-\hat{\rm H} |\psi_i^\prime \rangle\,=\,-\hat{\rm H} 
\left( \sum_m U_{im}^* |\psi_m \rangle \right)\;\;,
\end{equation}
\begin{equation}
\label{eq:g_prec_i} 
{\bf G}_i\,=\,\sum_n U^{* \dagger}_{in} {\bf G}^\prime_n\,=\,
-\,\hat{\rm H} |\psi_i \rangle \; \; . 
\end{equation}
Such preconditioned gradients ${\bf G}_i$ greatly improve the
convergence rate, given that higher bands are now updated with the same 
speed as lower filled bands, and are much cheaper to compute
than the ${\bf g}_i$ in Eq.~(\ref{eq:g_i}).
In addition to this occupancy preconditioning, a standard kinetic-energy
preconditioning should also be used in plane-wave calculations.
One iteration on the orbitals consists thus of several operations:
1) each preconditioned gradient $- \hat{\rm H} |\psi_i \rangle $ is calculated,
conjugated with the previous search direction, and
projected out of the subspace spanned by the orbitals
to assure first-order orthonormality along the search;
2) the first derivative of the free energy along the multidimensional 
(all bands, 
all plane-waves, and all ${\bf k}$-points) line is calculated, and
a trial step along the search line is taken;
3) after reorthogonalizing the orbitals, the new free energy 
provides the third constraint to identify the optimal, parabolic minimum
along the search line.

\begin{figure} 
\epsfxsize=3.3 truein
\epsfbox{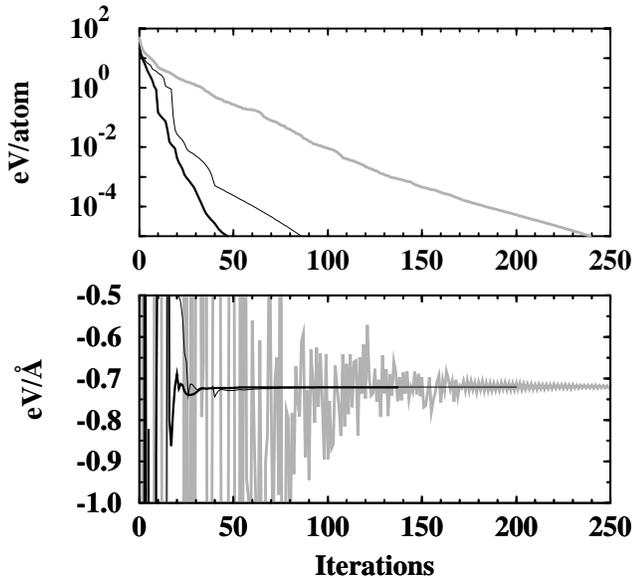}
\caption{Convergence of the total free energy (upper panel, 
semi-logarithmic scale) and of the force on the surface atom (lower 
panel) in the 15-layer Al(110) slab.  Grey line, ABV; thin
solid line, ensemble-DFT with 2 iterations in the inner loop; thick
solid line, ensemble-DFT with 4 iterations in the inner loop.}
\label{fig:conv_BW.ps} 
\end{figure} 

\begin{figure} 
\epsfxsize=3.3 truein
\epsfbox{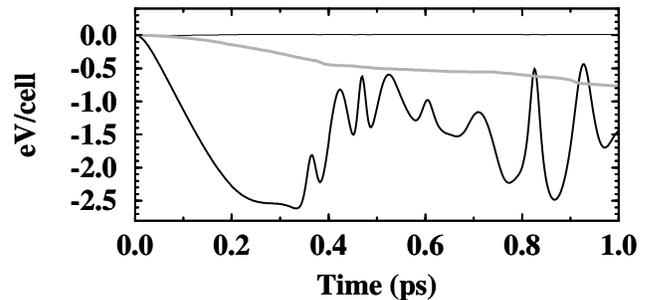}
\caption{Conservation of the constant of motion in a molecular-dynamics
run for the 15-layer Al(110) slab. The bottom curve is the
electronic free energy; the top two curves add to that
the kinetic energy of the ions (the grey line is for ABV,
the solid thin line is for ensemble-DFT with 2 iterations in the inner loop).}
\label{fig:const2.ps} 
\end{figure}

The complete algorithm provides a remarkably robust and efficient convergence.
As a paradigmatic case we present here results for a unit cell that
is 32 \AA\ long, and contains a
15-layer $1\!\!\times\!\!1$ Al(110) slab. We use the single
${\bf k}$-point 
$\frac{2\pi}{a_0} (\frac{1}{4},\frac{1}{4},\frac{1}{4})$,
a fictitious Gaussian temperature\cite{gillan2} of $4$ eV, and $64$ orbitals.
The large value of the temperature assures that the coarse 
sampling is sufficient; similar results are obtained with smaller
and more physical temperatures.
Fig.~\ref{fig:conv_BW.ps} monitors the convergence of the total
free energy and of the Hellmann-Feynman forces as a function
of the number of iterations on the $\{\psi_i\}$; we compare an 
optimal all-bands variational implementation (ABV)\cite{gillan2}
with the scheme that we have described (ensemble-DFT) \cite{cpu}.
The improved convergence for the total energies, and particularly 
for the Hellmann-Feynman forces, is clearly apparent.
It should be stressed that, at variance with the ABV case,
the behavior of the total free energy in the line
searches in both the outer and the inner loops is accurately parabolic;
interpolated minima are usually fractions of a percent off their true values.
The improved convergence of the ionic forces leads in particular to a
much tighter conservation of the constant of motion in molecular dynamics 
simulations. We show in Fig.~\ref{fig:const2.ps} the results of
a run for our Al(110) slab. The timestep is 2 fs;
the ions are moved after a fixed tolerance in the convergence of the 
free energy is reached (identical results are obtained if a 
fixed number of iterations is used).
The systematic drift of the constant of motion stabilizes 
after $\sim 0.3$ ps of thermalization to $-0.6$ eV/cell/ps for the
ABV case, and to $-0.0008$ eV/cell/ps for ensemble-DFT.
Such stability opens the way to inexpensive molecular dynamics 
simulation of large metallic systems even on common workstations\cite{nicola}.

N.\ M.\ would like to thank A.\ De Vita and J.\ White for many helpful
discussions. We gratefully acknowledge financial support from 
grant ERBCHBICT920192 (N.M.), NSF grant DMR-91-15342 (D.V.),
and access to the Hitachi S3600 located at Hitachi Europe's 
Maidenhead (UK) headquarters.


\begin{references}

\bibitem{dft-lda} See e.g.\ R.\ O.\ Jones and O.\ Gunnarsson,
Rev.\ Mod.\ Phys.\ {\bf 61}, 689 (1989).

\bibitem{cp-cg} R.\ Car and M.\ Parrinello, Phys.\ Rev.\ Lett.\
{\bf 55}, 2471 (1985); M.\ C.\ Payne, M.\ P.\ Teter, D.\ C.\ Allan,
T.\ A.\ Arias, and J.\ D.\ Joannopoulos, Rev.\ Mod.\ Phys.\ {\bf 64},
1045 (1992); D.\ Vanderbilt, Phys.\ Rev.\ B {\bf 41}, 7892 (1990).

\bibitem{smearing} C.-L.\ Fu and K.-M.\ Ho, Phys.\ Rev.\ B {\bf 28},
5480 (1983); M.\ J.\ Gillan, J.\ Phys.: Condens.\ Matter
{\bf 1}, 689 (1989).

\bibitem{mermin} N.\ D.\ Mermin, Phys.\ Rev.\ {\bf 137}, A1441 (1965).

\bibitem{genentropy} A.\ De Vita, Ph.D.\ Thesis, University of Keele (1992);
N.\ Marzari {\it et al.}, to be published.

\bibitem{alavi} A.\ Alavi, J.\ Kohanoff, M.\ Parrinello, and D. Frenkel,
Phys.\ Rev.\ Lett.\ {\bf 73}, 2599 (1994).

\bibitem{fqhe} O.\ Heinonen, M.\ I.\ Lubin, and M.\ D.\ Johnson,
Phys.\ Rev.\ Lett.\ {\bf 75}, 4110 (1995); M.\ Ferconi, M.\ R.\ Geller,
and G.\ Vignale, Phys.\ Rev.\ B {\bf 52}, 16357 (1995).

\bibitem{ensemble} See e.g.\
R.\ G.\ Parr and W. Yang, {\it Density-Functional Theory of
Atoms and Molecules}, Oxford University Press, New York (1989).

\bibitem{apj} T.\ A.\ Arias, M.\ C.\ Payne, and J.\ D.\ Joannopoulos,
Phys.\ Rev.\ Lett.\ {\bf 69}, 1077 (1992).

\bibitem{nicola} N.\ Marzari, Ph.D.\ Thesis, University of
Cambridge (1996); N.\ Marzari {\it et al.}, to be published.

\bibitem{gillan} M.\ J.\ Gillan, private communication.

\bibitem{prec} The mixing factor used to conjugate a direction
depends on the scalar product between the original and the
preconditioned gradient; we just use the preconditioned gradient on both 
sides, without noticing a deterioration in the convergence.

\bibitem{gillan2} J.\ M.\ Holender, M.\ J.\ Gillan, M.\ C.\ Payne,
and A.\ D.\ Simpson, Phys.\ Rev.\ B {\bf 52} 967 (1995).

\bibitem{cpu} The cost of one ensemble-DFT iteration
(with 2 $f_{ij}$ iterations in the inner loop) ranges between 
$2/3$ and $2$ (in ABV units); the extrema are for the 
Fourier-transform limited regime
with or without the $\{\psi_i\}$ stored in real-space.
The cost in the cubic-scaling regime is $\sim 3/2$, or lower if 
the non-local pseudopotentials are in the reciprocal-space 
representation.
\end{references}
\end{document}